\documentclass{aastex}
\usepackage{emulateapj5}
\usepackage{apjfonts}

\newcommand{\abs}[1]{\ensuremath{\left| \mathbf{#1} \right|}}

\begin{document}

\slugcomment{Published in the Astrophysical Journal, 568, 38}

\title{DASI First Results: A Measurement of the Cosmic Microwave Background Angular Power Spectrum}

\author{N.\ W.\ Halverson, E.\ M.\ Leitch, C.\ Pryke,
J.\ Kovac, J.\ E.\ Carlstrom}
\affil{University of Chicago,
5640 South Ellis Ave.,
Chicago, IL 60637}
\author{W.\ L.\ Holzapfel}
\affil{University of California,
426 Le Conte Hall,
Berkeley, CA 94720}
\author{M.\ Dragovan}
\affil{Jet Propulsion Laboratory,
California Institute of Technology,
4800 Oak Grove Drive,
Pasadena, CA 91109}
\author{J.\ K.\ Cartwright, B.\ S.\ Mason, S.\ Padin,
T.\ J.\ Pearson, A.\ C.\ S.\ Readhead, M.\ C.\ Shepherd}
\affil{California Institute of Technology,
1200 East California Boulevard,
Pasadena, CA 91125}

\begin{abstract}
We present measurements of anisotropy in the Cosmic Microwave
Background (CMB) from the first season of observations with the Degree
Angular Scale Interferometer (DASI). The instrument was deployed at
the South Pole in the austral summer 1999--2000, and made observations
throughout the following austral winter. We present a measurement of
the CMB angular power spectrum in the range $100 < l < 900$ in
nine bands with fractional uncertainties in the range 10--20\% and
dominated by sample variance. In this paper we review the formalism
used in the analysis, in particular the use of constraint matrices to
project out contaminants such as ground and point source signals and
to test for correlations with diffuse foreground templates. We find no
evidence of foregrounds other than point sources in the data, and find
a maximum likelihood temperature spectral index $ \beta = -0.1 \pm
0.2$ (1\,$\sigma$), consistent with CMB. We detect a first peak in the
power spectrum at $l \sim 200$, in agreement with previous
experiments. In addition, we detect a peak in the power spectrum at $l
\sim 550$ and power of similar magnitude at $l \sim 800$ which are
consistent with the second and third harmonic peaks predicted by
adiabatic inflationary cosmological models.

\end{abstract}


\section{Introduction}
\label{sec:intro}

Subtle temperature fluctuations in the Cosmic Microwave Background
(CMB) radiation, first observed by the COBE DMR experiment
\markcite{smoot92}(Smoot {et~al.} 1992), offer a glimpse of the early Universe at the epoch of
matter-radiation decoupling, well before the non-linear gravitational
collapse of matter led to the structure we see in the Universe today.
The theory of inflation, originally proposed as an explanation for
apparent isotropy in the CMB and a spatial curvature near unity
\markcite{guth81,linde82,albrecht82}({Guth} 1981; Linde 1982; Albrecht \& Steinhardt 1982), was further developed to predict
nearly scale-invariant adiabatic density perturbations arising from
quantum fluctuations as the source of inhomogeneity in the Universe
\markcite{guth82,hawking82,starobinskii82,bardeen83}({Guth} \& {Pi} 1982; {Hawking} 1982; Starobinsky 1982; {Bardeen}, {Steinhardt}, \&  {Turner} 1983). On angular scales
causally disconnected at the epoch of decoupling ($\gtrsim 1\degr$),
anisotropy in the CMB reflects the primordial inhomogeneity of matter
\markcite{white94}(see, e.g., White, Scott, \& Silk 1994). On scales smaller than the sound
horizon at the time of decoupling, the primordial adiabatic density
perturbations force gravity-driven acoustic oscillations in the
photon-baryon fluid which lead to a harmonic series of peaks in the
CMB angular power spectrum \markcite{bond84,vittorio84}(Bond \& Efstathiou 1984; {Vittorio} \& {Silk} 1984). A detection of a
series of peaks in the CMB angular power spectrum would provide strong
evidence for the inflationary view of the early Universe, and would make
alternate theories of structure formation such as defect models
difficult to support \markcite{albrecht96}(Albrecht {et~al.} 1996). A third harmonic peak of
comparable or greater magnitude than the second is the signature of
baryonic matter re-compressing in dark matter potential wells --- a
detection of significant power in the third peak region would strongly
support the presence of dark matter in the Universe
\markcite{hu96a}(Hu \& White 1996). Within the context of standard cosmological models, the
observed CMB angular power spectrum can be used to determine
fundamental cosmological parameters \markcite{knox95,jungman96}({Knox} 1995; {Jungman} {et~al.} 1996).

The angular scale of the first acoustic peak, coupled with the sound
horizon size at last scattering, provides a classical angular-diameter
distance measure of spatial curvature.  Recent experiments have
determined the location of the degree-scale first peak, providing
strong evidence for a spatially flat Universe
\markcite{miller99,mauskopf00a,debernardis00,hanany00}({Miller} {et~al.} 1999; {Mauskopf} {et~al.} 2000; {de Bernardis} {et~al.} 2000; {Hanany} {et~al.} 2000). Additional
parameters such as the baryonic matter density ($\Omega_b\,h^2$) can
be extracted by resolving the higher-order peaks in the power
spectrum.  Hints of structure appear to be present in some smaller
angular scale CMB power spectra \markcite{hanany00,padin01}({Hanany} {et~al.} 2000; {Padin} {et~al.} 2001a), but there
has been no clear detection of the predicted higher-order acoustic
peaks.

Until recently, most observations of degree-scale CMB anisotropy have
been made with ground and balloon based single-dish experiments.  The
Degree Angular Scale Interferometer (DASI), like its sister instrument
the CBI \markcite{pearson00}({Pearson} {et~al.} 2000) and the VSA \markcite{jones96}({Jones} 1996), is a new
compact interferometer constructed specifically for observations of
the CMB. Since interferometers are inherently differencing
instruments, they allow measurements of anisotropy in the
CMB that are free of many of the systematics which need to be controlled in
non-interferometric CMB experiments. DASI is designed to measure the
angular power spectrum of the CMB and produce high signal-to-noise
images on angular scales corresponding to the first three acoustic
peaks predicted by adiabatic inflationary models for a flat Universe.

In this, the second of three papers describing the results of the
first season of DASI observations, we focus on the determination of
the CMB power spectrum from the calibrated data. This paper includes a
discussion of the data analysis method, potential sources of
astronomical and terrestrial contamination in the data, tests of data
consistency, and the resulting angular power spectrum. Detailed
descriptions of the instrument, observations, and data calibration 
are given in \markcite{leitch01}Leitch {et~al.} (2001,  hereafter Paper I). The
extraction of the cosmological parameters from the DASI angular power
spectrum is presented in a third paper, \markcite{pryke01}Pryke {et~al.} (2001,  hereafter Paper
III).

\section{Instrument}
\label{sec:inst}
An extensive discussion of the instrument design and its performance
are given in Paper I; here we provide a brief overview stressing the
aspects of the instrument that are particularly relevant to the
measurement of the CMB angular power spectrum. 

Interferometers offer several unique features which make them well
suited for sensitive measurements of CMB anisotropy: 1) the Fourier
transform of the sky plane is measured directly; the effective sky
brightness differencing is instantaneous, 2) $180\degr$ phase
switching at the receivers with synchronized post-detection
demodulation is used to reduce instrumental offsets (DASI uses both
fast (40~KHz) phase switching with hardware demodulation and slow
($\sim 1$ Hz) phase switching with software demodulation to reduce
offsets to well below the $\mu$K level), and 3) the effective window
function in $l$-space is well understood and uncertainties in the
primary beam response do not lead to uncertainties in the resultant
power spectrum that become large at small angular scales.

An interferometer directly samples the Fourier transform of the sky
brightness distribution. The response pattern on the sky for a given
pair of antennas is a sinusoidal fringe pattern attenuated by the
primary beam of the individual antennas. For a pair of antennas with a
physical separation (baseline) vector $\mathbf{b}$, the center of the
measured Fourier wavevector components, labeled $\mathbf{u}$ or
$(u,v)$, is given by $u = b_x/\lambda$ and $v = b_y/\lambda$, where
$b_x$ and $b_y$ are the projections of the baseline normal to the line
of sight, and $\lambda$ is the observing wavelength. The approximate
conversion to multipole moment is given by $l \approx
2\pi\left|\mathbf{u}\right|$ \markcite{white99a}(White {et~al.} 1999a), with a width $\Delta l$
that is related to the diameter of the apertures in units of the observing
wavelength (see \S\ref{sec:formalism}).

Both real and imaginary components of the Fourier plane can be
measured with a complex correlator. The real (even) component is
measured by correlating the pair of signals without a relative
phase delay; the imaginary (odd) component is measured by correlating
the signals with a $90\degr$ phase shift introduced into one of the
signal paths. The averaged correlated output of the interferometer is
called the \emph{visibility} (see eq.~[\ref{eqn:vis}]) and is the
fundamental data product.

DASI is a 13-element interferometer operating at 26--36~GHz. 
The 13 antenna elements are arranged in a threefold
symmetric configuration on a common mount which can point in
azimuth and elevation. The mount is also able to rotate the array of
horns about the line of sight to provide additional $(u,v)$ coverage as
well as the ability to perform consistency checks.  
The 13 elements provide 78 baselines with baseline lengths in the range
25--121~cm. 
The configuration of the horns was chosen to provide dense
coverage of the CMB angular power spectrum from $100 < l < 900$. 

Each DASI antenna consists of a 20-cm aperture-diameter lensed
corrugated horn which defines the $\sim 3 \fdg 4$ FWHM field of view
of the instrument.  The receivers use cooled low-noise high electron
mobility transistor (HEMT) amplifiers \markcite{pospieszalski95}(Pospiezalski {et~al.} 1995), and
have system noise temperatures referred to above the atmosphere ranging
from 18~K to 35~K at the center of the band.  The receivers
downconvert the 26--36~GHz RF band to a 2--12~GHz IF band. Each
receiver IF is further split and downconverted to ten 1~GHz wide bands
centered at 1.5 GHz.  An analog correlator \markcite{padin00}({Padin} {et~al.} 2001b) processes
the 1 GHz bands into 780 complex visibilities.

The stability of the instrument, its location at the South Pole, and
the fact that its mount is fully steerable, have given us great
flexibility in designing and adapting our observing strategy. We are
able to choose fields to avoid foreground contamination, balance
sensitivity and sky coverage, and observe in patterns that reject
ground and other spurious signals while producing datasets containing
correlations which are computationally tractable.

\section{Observations}
\label{sec:obs}

CMB fields were observed during the period spanning 05 May--07
November 2000.  The data presented here comprise 97 days of
observation, representing an observing efficiency of better than 85\%
(of the days devoted exclusively to CMB observations), with the
remainder lost to hardware maintenance and repairs. Observations were never
prevented due to weather, and only 5\% of data were lost due to 
weather based edits, confirming previous assessments of the exceptional 
quality of the site \markcite{lay98,chamberlin97}({Lay} \& {Halverson} 2000; Chamberlin, Lane, \& Stark 1997).  

The presence of near-field ground contamination at a level well above
the CMB signal limits our ability to track single fields over a wide
range in azimuth.  Repeated tracks over the full azimuth range show a
strong variation of the ground with direction, with amplitudes of tens
of Jy on some of the shortest baselines, but there is little
evidence for time variability on periods as long as five days.  One
advantage of observing near the South Pole is that sources track at a
constant elevation, which enables us to observe several sources at
constant elevation over a given range in azimuth.  Observations were
divided among 4 constant declination (elevation) rows of 8 fields, on
a regular hexagonal grid spaced by 1h in right ascension, and $6\degr$
in declination. The grid center was selected to avoid the Galactic
plane and to coincide with a global minimum in the IRAS 100~$/mu$m map
of the southern sky. Each field in a row was observed over the same
azimuth range, leading to a nearly identical ground contribution.  The
elevation of the rows are $61\degr, 67\degr, 55\degr, 49\degr$, which
we label the A, B, C and D rows for the order in which they were
observed (see Paper I for full coordinates).  The field separation of
1h in RA represents a compromise between immunity to time
variability of the ground signal and a desire to minimize inter-field
correlations.

A given field row was observed daily over two azimuth ranges, for a
total of 16 hours per day, with the remainder of the time divided
among various calibration and pointing tasks (see Paper I).  Phase and
amplitude calibration were accomplished through observations of bright
Galactic sources, permitting determination of the calibrator flux on
all baselines to better than 2\%. Absolute pointing error determined
by offsets between DASI detected point source positions and PMN
southern catalog coordinates \markcite{wright94}({Wright} {et~al.} 1994) was less than
$2\arcmin$, with a drift $\ll 1\arcmin$ over the period during which
each row was observed. The number of days for which each of the four rows was
observed is 14, 24, 28 and 31 for the A, B, C and D rows,
respectively, for a total integration time of 28--62 hours per field.

\section{Calibration \& Data Reduction}
\label{sec:reduc}

Absolute calibration of the telescope was achieved through
measurements of external thermal loads; the calibrations were then
transferred to bright astronomical sources.  The flux scales resulting
from two independent calibrations performed in February 2000 and
February 2001 are found to agree to 0.3\%, consistent with our
estimate of 1\% overall statistical uncertainty in the measurement and
transfer procedure.  The systematic uncertainty in determining load
coupling and effective temperature is 3\%, which is the dominant
contribution to the uncertainty in our overall flux scale. This
uncertainty, expressed as a percentage of $C_l$, is 7\% at 1\,$\sigma$
and is constant across all power spectrum bands. Band-power
measurements are also affected, though weakly, by errors in the
estimated aperture efficiency, on which our uncertainty is $4\%$ (see
Paper I).  This contributes a band-power uncertainty which is constant
at $4\%$ except in the three lowest-$l$ bands, where a cancellation of
errors causes it to decrease.  In using the current DASI results for
parameter estimation (Paper III), we have found no significant
difference between treating this small beam uncertainty separately
with its low-$l$ variation included, and folding it together with the
$l$-independent flux scale uncertainty. We therefore adopt a total
combined calibration uncertainty of 8\% (1\,$\sigma$), expressed as a
fractional uncertainty on the $C_l$ band powers (4\% in $\Delta T/T$).

Raw data from the correlators, along with monitoring data from various
telescope systems, are accumulated in 8.4-s integrations.
These short integrations are edited before being
combined for analysis. Baselines are rejected for which the phase
offset or relative gain between the real and imaginary multipliers
exceed nominal values. Data are also rejected when an
LO has lost phase lock, when a receiver has warmed, or to trim field
scans so that all eight fields are observed over precisely the same
azimuth range. We also edit data for which noise correlations 
between baselines indicate strong atmospheric fluctuations.  

The edited and calibrated data are combined into 1-hr bins, with
uncertainty in the bins estimated from the sample variance of the
8.4-s integrations. In order to implement ground contamination common
mode rejection, it is necessary that a given visibility be measured
for all 8 fields in a row; we cut all baselines that do not satisfy
this criterion.  We apply more stringent edits for $(u,v)$ radii $<
40$, which we find are more susceptible to contamination.  For these
visibilities, we retain only data for which both the sun and moon are
below the horizon. To minimize the risk of biasing the power spectrum
results, we do not edit the data based on the level of the signal. We
have varied the threshold values of the weather, calibrator, and
lunar/solar edit criteria with no significant effect on the
results. Collectively, these edits reject about 40\% of the data. See
Paper I for a more comprehensive description of the data edits.

All observations of a given set of fields are then combined, and it is
these 1560 combined visibilities per field (78 complex baselines $\times$ 10
correlator channels, before edits) which form the input to the angular
power spectrum likelihood analysis.

\section{Analysis}
\label{sec:analysis}

\subsection{Formalism}
\label{sec:formalism}

The DASI instrument makes direct measurements of the Fourier
plane, and the angular power spectrum can be extracted
from the data without creating an image. The calibrated output of
the interferometer is the visibility,
\begin{equation}
V(\mathbf{u}) =  \frac{2k_BT}{\lambda^2}g(\lambda)\widetilde{A}(\mathbf{u},\lambda) 
	\ast \frac{\widetilde{\Delta T}}{T}(\mathbf{u}),
\label{eqn:vis}
\end{equation}
which is the convolution of the Fourier Transform of the sky
brightness distribution, $\widetilde{\Delta T}(\mathbf{u})/T$, with
the antenna aperture field autocorrelation function,
$\widetilde{A}(\mathbf{u},\lambda)$, and $g(\lambda)$ is a $\sim2$\%
correction between the Rayleigh-Jeans and Planck functions. The
aperture field autocorrelation function
$\widetilde{A}(\mathbf{u},\lambda)$ is radially symmetric,
peaking at $\abs{u} = 0$, and tapering smoothly to zero at
$\abs{u} = D/\lambda$, where $D$ is the aperture diameter. In the flat-sky 
limit, which is appropriate for the $\sim 3\fdg4$ FWHM DASI fields,
\begin{equation}
S(\abs{u}) \equiv \left<\left|\frac{\widetilde{\Delta T}}{T}(\mathbf{u})\right|^2\right> 
 	\simeq C_l\Big|_{l = 2 \pi \abs{u}} \mbox{ for}\ \abs{u} \gtrsim 10
\end{equation}
\markcite{white99a,hobson95}(White {et~al.} 1999a; Hobson, Lasenby, \& Jones 1995); we assume $l = 2 \pi \abs{u}$ over the
$l$-range to which DASI is sensitive. For a single visibility, a
simple quadratic estimator $\hat \mathcal{S}$ of the quantity $2 \pi
\abs{u}^2 S(\abs{u}) \approx l(l+l)C_l/(2 \pi)$ is given by
\markcite{white99b}(White {et~al.} 1999b)
\begin{eqnarray}
\hat \mathcal{S} & = & 2 \pi \abs{u}^2 \frac{\left|V(\mathbf{u})\right|^2 - N}
	{(2k_BT / \lambda^2)^2 g(\lambda)^2 \int{ d\mathbf{u'}
	\widetilde{A}(\mathbf{u'},\lambda)^2}} \\
	& = & \left(\frac{17.4\ \mu \mathrm{K}^2}{\mathrm{Jy}^2}\right)
		\abs{u}^2 (\left|V(\mathbf{u})\right|^2 - N) 
\label{eqn:sqedasi}
\end{eqnarray} 
where $N$ is the instrument noise variance for the measured visibility
$V(\mathbf{u})$ and equation~(\ref{eqn:sqedasi}) gives the number specific to
the DASI apertures and a power spectrum in units of $\mu
\mathrm{K}^2$. The variance of the visibility is thus directly related
to $C_l$ centered at the baseline length $\left|\mathbf{u}\right| =
l/(2 \pi)$, with width $\Delta \abs{u} \simeq 12$ (FWHM) determined by the
width of the aperture field autocorrelation function.

While the simple quadratic estimator above is useful for understanding
the relationship between the visibility and the angular power
spectrum, we have chosen a maximum likelihood method in the present
analysis. We have adopted the iterated quadratic estimator approach
of \markcite{bond97}{Bond}, {Jaffe}, \& {Knox} (1998) to find the maximum likelihood values of the angular
power spectrum for a piecewise flat $l(l+1)C_l/(2\pi)$ power spectrum in nine
bands. A data vector $\Delta$ of length $N = 1560 \times 32$ (before
data edits) is constructed by combining observations of each
visibility for each of the 32 fields. The likelihood function for a
set of parameters $\mathbf{\kappa}$ is
\begin{equation}
\mathcal{L}_\mathbf{\Delta}(\mathbf{\kappa}) = \frac{1}{(2\pi)^{N/2}
	\left|C(\mathbf{\kappa})\right|^{1/2}} \exp{(-\onehalf
	\mathbf{\Delta}^T C(\mathbf{\kappa})^{-1} \mathbf{\Delta})},
\end{equation}
where the covariance matrix
\begin{equation}
C(\mathbf{\kappa}) = C_T(\mathbf{\kappa}) + C_n + C_C
\end{equation}
is the sum of the theory, noise, and constraint covariance matrices,
described below, and is a function of the parameters. The parameters
$\mathbf{\kappa}$ which we estimate are the band powers,
$2 \pi u^2S(u)\approx l(l+1)C_l/(2\pi)$.  The theory covariance matrix,
$C_T$, is given by
\begin{equation}
C_T(\mathbf{\kappa}) \equiv \left<\mathbf{V}\mathbf{V}^T\right> = \sum_p \mathbf{\kappa}_p B_p,
\end{equation}
where $\mathbf{V}$ is the vector of noiseless theoretical visibilities
and the sum is over the piecewise flat bands $p$. The matrices $B_p$
represent the instrument filter functions to fluctuation power on the
sky. They are constructed from the overlap integral of the aperture
field autocorrelation functions of pairs of baselines where they
sample the same Fourier modes on the sky,
\begin{eqnarray}
{B_p}_{ij} & = & \frac{1}{2\pi}\frac{(2k_BT)^2}{\lambda_i^2 \lambda_j^2} 
	g(\lambda_i)g(\lambda_j) \nonumber\\
	& & {} \times \frac{1}{2} \int_{\left|\mathbf{u}_1\right|_p}
	^{\left|\mathbf{u}_2\right|_p}
	{\frac{d\mathbf{u'}}{\left|\mathbf{u}'\right|^2}\,
	\widetilde{A}(\mathbf{u}_i-\mathbf{u'},\lambda_i)
        [\widetilde{A}(\mathbf{u}_j-\mathbf{u'},\lambda_j)} \nonumber\\ 
	& & \hspace{1.55in} \pm \widetilde{A}(\mathbf{u}_j+\mathbf{u'},\lambda_j)].
\label{eqn:bpij}
\end{eqnarray}
Here $\left|\mathbf{u}_1\right|_p,\left|\mathbf{u}_2\right|_p$ are the
lower and upper radial limits of band $p$, and $+$ and $-$ are used
for the real and imaginary parts of the visibility, respectively,
which we treat as separate elements in the data vector.  The vectors
$\mathbf{u}_i$ and $\mathbf{u}_j$ are the $(u,v)$ coordinates of the
visibility data vector elements $\Delta_i$ and $\Delta_j$ respectively
\markcite{white99a,white99b,hobson95}(White {et~al.} 1999a, 1999b; Hobson {et~al.} 1995). Equation~(\ref{eqn:bpij}) is simply
an overlap integral between an aperture autocorrelation function
centered at $\mathbf{u}_i$ and ones centered at $\pm\mathbf{u}_j$;
only one of the two overlap terms is non-zero, except for the shortest
baselines, and the total integral is non-zero only when $\mathbf{u}_i
\pm \mathbf{u}_j \leq D/\lambda_i + D/\lambda_j$, i.e., when the two
visibilities are sensitive to some of the same wavevector components
in the $(u,v)$ plane. We use the theoretical aperture fields in this
calculation, which is justified by the good agreement between the
theoretical and measured beams \markcite{halverson01b}(see Paper I and
 Halverson \& Carlstrom 2001). Fields are separated such that the inter-field data
vector elements are essentially uncorrelated except in the highest
elevation row, for which we calculate the appropriate correlations.

The noise covariance matrix of the combined visibility data vector is
diagonal, with elements ${C_n}_{ii} = \sigma^2_{ii}$ estimated from
the sample variance in the 8.4-s integrations over the 1-hr
observations. To verify the assumption that $C_n$ is diagonal, we have
calculated the sample covariance matrix from the data in each of the
ten frequency channels for all 1-hr observations. We find rare
occasions where the visibilities are strongly correlated due to
atmospheric fluctuations. Our weather edits consist of cuts based on
the strength of these correlations; we cut observations in which any
off-diagonal correlation coefficient exceeds $\pm 0.36$, but the 
data consistency does not depend strongly on this value.

To reduce near-field ground contamination and point source
contributions to the power spectrum, we employ the constraint matrix
formalism described in \markcite{bond97}{Bond} {et~al.} (1998) to marginalize over potentially
contaminated modes in the data. Specifically, for a given mode $q$, we construct
a constraint matrix from the outer product of a template vector
$\mathbf{T_q}$,
\begin{equation}
C_{Cq} = \mathbf{T_q}\mathbf{T_q}^T
\end{equation}
and 
\begin{equation}
C_C = \alpha \sum_q C_{Cq},
\end{equation}
where $\alpha$ is a number large enough to de-weight the undesired
modes without causing the covariance matrix to become singular to
working precision. In practice we have found that $\alpha$ can range
over at least seven orders of magnitude without affecting the
results. As an example of a template vector, in the sub-space of the
data vector consisting of a single visibility observed in eight
fields, a template vector $\mathbf{T_q} = [1 1 1 1 1 1 1 1]^T$ is used
to constrain a common mode with the same amplitude in all eight fields
during any given 8-hr period of observation. This effectively rejects
a constant amplitude component, even one which varies in amplitude
between subsequent 8-hr observations, such as a temporally drifting
noise component with a period of many days. Any mode in the data which
can be described as a relative amplitude between data vector elements,
as in the example above, can be constrained. We use this method to
reduce near-field ground contamination in the field rows
(see \S\ref{sec:groundconstraints}) and contributions from point sources
with known positions (see \S\ref{sec:pointsourceconstraints}). It is
equivalent to marginalizing over these modes, with no knowledge of
their amplitude scale. For each point source we also use constraint
matrices to marginalize over arbitrary spectral indices, which we
approximate as an amplitude slope across the ten frequency channels. 
We emphasize that we do \emph{not} subtract ground components or point
sources from the data. Instead we render the analysis insensitive to
these modes in the data using the methods described above.

The covariance matrix is block diagonal by field row, which allows us
to invert the four sub-matrices in parallel. We further compress the
matrix by combining visibilities and covariance matrix elements from
adjacent frequency channels, which are nearby in the $(u,v)$ plane and
are therefore highly correlated. This data compression has the effect
of slightly increasing the uncertainties and the anti-correlations
between the resulting band-power estimates, but preserves the original
piecewise-flat theoretical power spectrum model.

The likelihood analysis software was extensively tested through
analysis of simulated data. The analysis software and data simulation
software were written by different authors in order to check for
potential errors in the analysis. Omitting the constraint matrix
leaves a sparse covariance matrix which can be rapidly inverted, and
we can analyze a simulated data set in a few minutes of CPU
time. Multiple simulated datasets were generated from an input model
power spectrum, each with independent sky and instrument noise
realizations; the analysis software recovered the input model power
spectrum within the estimated uncertainties, and these uncertainties
were found to match, on average, the scatter in the band-power
estimates over many realizations of the simulated data vector. We
found no evidence of bias in the maximum likelihood band-power
estimators. Ground signal and point source constraints were tested by
constraining these modes in simulated data which contained both ground
and point source components; both components are effectively
eliminated, and the constraint matrices do not introduce artifacts
into the power spectrum.

\subsection{Ground Constraints}
\label{sec:groundconstraints}

To remove sensitivity to the ground signal, we apply a constraint
which marginalizes over a common component across eight fields for
each visibility, as described above. Additionally, using sensitive
consistency tests described in \S\ref{sec:consistency}, we find
evidence of a temporally drifting component of the ground signal on 1-
to 8-hr time scales, subtle but present for all baselines and
noticeably stronger for short baselines. We therefore apply a linear
drift constraint to all visibilities, and a quadratic constraint for
${\left| \mathbf{u} \right|} < 40$. The additional constraints have
little effect on the power spectrum, which makes us confident that
sensitivity to ground signal is effectively eliminated.

\subsection{Point Source Constraints}
\label{sec:pointsourceconstraints}

As predicted for our experimental configuration \markcite{tegmark96}(Tegmark \& Efstathiou 1996),
point sources are the dominant foreground in the DASI data. To remove
point source flux contributions using the constraint matrix formalism
above, we require only the positions of the sources, {\it not their
flux densities}.  We constrain 28 point sources detected in the DASI data
itself, in which we can detect an 40~mJy source at beam center with $>
4.5\,\sigma$ significance. The estimated point source flux densities range
from 80~mJy to 7.0~Jy. We also constrain point sources from the PMN
southern (PMNS) catalog \markcite{wright94}({Wright} {et~al.} 1994) with 4.85~GHz flux densities,
$S_5$, which exceed 50~mJy when multiplied by the DASI primary
beam. We use this flux density limit for the constrained point sources because
the loss of degrees of freedom resulting from the inclusion of all
point sources in the PMNS catalog would be prohibitively large. We
have tested for the effect of possible absolute pointing error by
displacing the point source position templates. A uniform displacement
of the PMNS catalog coordinates by less than our estimated pointing error
of $2\arcmin$ (see Paper I) does not have a significant effect on the
angular power spectrum, except in the three highest-$l$ bins where the
effect is $\sim 10\%$. For the brightest point sources,
positions accurate to $<1\arcmin$ are required. We can extract
positions to the necessary accuracy from the DASI data (see Paper I).

In addition to the point sources constrained above, we make a
statistical correction for residual point sources which are too faint
to be detected by DASI or included in our PMN source table. To do
this, we estimate the point source number count per unit flux density
at 4.85~GHz, $dN/dS_5$, derived from the PMNS catalog, and the
distribution of 31~GHz to 4.85~GHz flux density ratios, $S_{31}/S_5$,
derived from new observations for this purpose with the OVRO 40~m
telescope in Ka band (paper in preparation). We proceed to calculate
the statistical correction for unconstrained residual point sources
$S_{31} > 1$~mJy using Monte Carlo techniques; we generate random
point source distributions at 4.85~GHz using $dN/dS_5$ and
statistically extrapolate the flux density of each source to each of
our ten frequency channels using $S_{31}/S_5$. These
simulated point sources are superimposed on CMB temperature
fluctuations and observed with DASI simulation software; a power
spectrum is then generated with the analysis software. The resulting
mean amplitudes and uncertainties of the residual point source
contribution to the nine band powers are [$20 \pm 70$, $70 \pm 80$,
$90 \pm 70$, $180 \pm 70$, $240 \pm 80$, $330 \pm 100$, $400 \pm 100$,
$500 \pm 170$, $430 \pm 170$] $\mu K^2$. The reported uncertainties
are due to sky sample variance of the point source population in the
simulations, uncertainty in $dN/dS_5$, and uncertainty in
$S_{31}/S_5$. The residual point source contribution diminishes in the
ninth band since that band power is dominated by visibilities from the
highest frequency channels where the average point source flux density
is lower relative to its mean flux density across all ten frequency
channels. We use these statistically estimated amplitudes and
uncertainties to adjust our CMB band-power estimates and uncertainties
reported below.

\section{Results}
\label{sec:results}

\begin{figure*}[t]
\epsscale{1.5}
\plotone{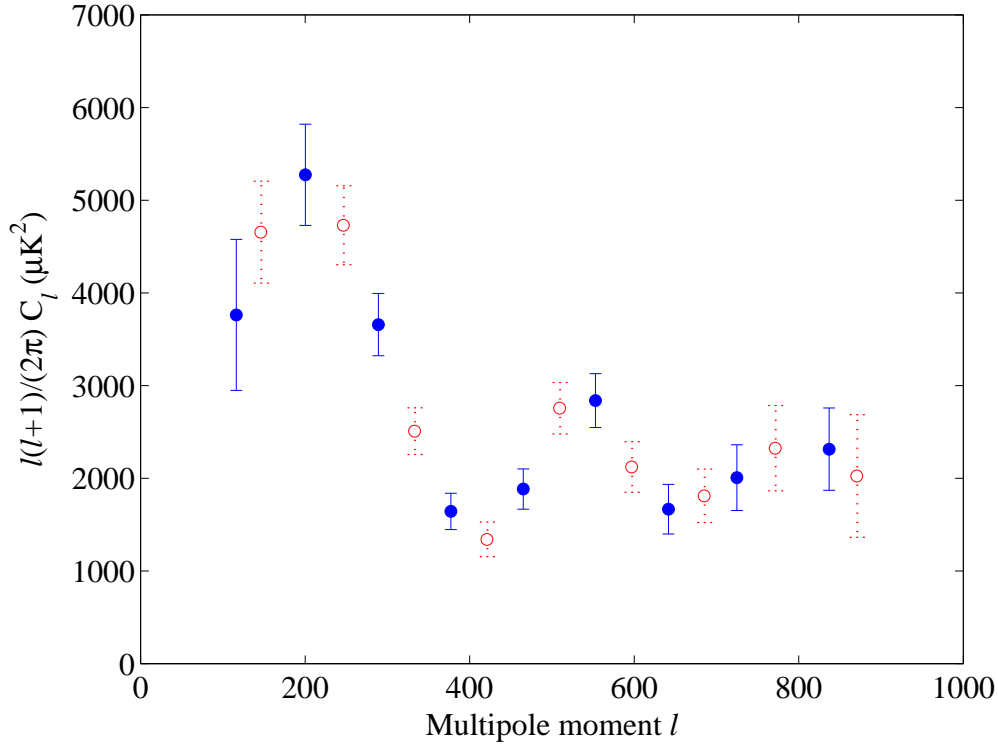}
\caption{\label{fig:ps} The angular power spectrum from the first
season of DASI observations, plotted in nine bands (filled
circles). We have analyzed the same data in nine bands shifted to the
right (open circles). The alternate set of band powers are shown to
demonstrate the robustness of the likelihood analysis procedure. To extract
cosmological parameters (Paper III), only the nine bands shown in the
primary analysis (filled circles) are used. Adjacent bands are
anticorrelated at the 20\% level (see Table \ref{tab:corr}). In
addition to the uncertainties plotted above, there is a calibration
uncertainty of 8\%, expressed as a fractional uncertainty on the $C_l$
band powers (4\% in $\Delta T/T$), which is completely correlated
across all bands due to the combined flux scale and beam
uncertainties.}
\end{figure*}

The CMB angular power spectrum from the first season of DASI data is
shown in Figure~\ref{fig:ps}, with maximum likelihood estimates of
nine band powers, piecewise flat in $l(l+1)C_l/(2\pi)$, spanning the
range $l = $100--900.  Adjacent bands are anticorrelated at the 20\%
level. In addition, we show an alternate analysis of the same data,
for nine bands shifted to the right with respect to the original band
edges, in order to demonstrate the robustness of the analysis against
possible effects due to the anticorrelation of adjacent bands. Note
that these two analyses use the same data to estimate band powers in
two different piecewise-flat theoretical power spectra; only the first
nine-band analysis (filled circles) is used for the cosmological
parameter estimation described in Paper III. While increasing the
number of bands above nine may in principle provide more information
about the underlying power spectrum, we have found that this does not
significantly improve our ability to constrain cosmological parameters
(see Paper III).

In a separate analysis, we fit for the maximum likelihood value of an
additional parameter, the temperature spectral index of the
fluctuations, $\beta$, where $T \propto \nu^{\beta}$. Fitting a single
spectral index for all nine bands, we find $ \beta = -0.1
\pm 0.2$ (1\,$\sigma$), while fitting a separate spectral index
for $l < 500$ and $l > 500$  yields $\beta = -0.2 \pm 0.3$ and
$0.0 \pm 0.4$ respectively, indicating the fluctuation power is
consistent with CMB.

Values and marginal uncertainties for the angular power spectrum
in the primary nine bands are given in Table~\ref{tab:ps}. The center
and $e^{-1/2}$ widths of the bands are calculated using
band-power window functions adapted from \markcite{knox99}{Knox} (1999) which are
plotted in Paper III. These are the relevant window functions for
calculating the expectation value of the band power given a
theoretical power spectrum. We give the ratio of the uncertainty due
to sky sample variance to the uncertainty due to noise,
$\sigma_{\mathrm{s}} / \sigma_{\mathrm{n}}$, estimated using the
offset log-normal formalism of \markcite{bond98}{Bond}, {Jaffe}, \& {Knox} (2000). In their notation,
$\sigma_{\mathrm{s}} / \sigma_{\mathrm{n}}$ is given by
$\mathcal{C}_B/x_B$, where $\mathcal{C}_B$ is the band power estimate 
expressed as $l(l+1)C_l/(2\pi)$ and $x_B$ is proportional to the instrument
noise contribution to the band-power uncertainty. These values may be
used to estimate the non-Gaussianity in the band-power
marginal likelihood distributions for parameter estimation calculations ---
asymmetric uncertainties due to non-Gaussianity are negligible for
most of our band powers and we do not plot them here. We also tabulate
the band-power correlation matrix (Table \ref{tab:corr}).  All of the
data products necessary for performing cosmological parameter
estimation from this data are available at our website\footnote{\tt
http://astro.uchicago.edu/dasi}.

\begin{table*}[t]
\caption{\label{tab:ps}Angular power spectrum band powers and uncertainties}
\begin{center}
\begin{tabular}{cccccr}\hline\hline

\rule [-3mm]{0mm}{8mm}
$l_\mathrm{eff}$\tablenotemark{a} & $l_\mathrm{l}$\tablenotemark{b} 
& $l_\mathrm{h}$\tablenotemark{b} & $l(l+1)C_l/(2\pi)\ (\mu K^2)$ & 
$\sigma_{\mathrm{s}}/\sigma_{\mathrm{n}}$ \tablenotemark{c}
& $l(l+1)C_l^\mathrm{E1-E2}/(2\pi)\ (\mu K^2)$\tablenotemark{d}\\
\hline
118     &       104     &       167     &       $3770 \pm 820$  &       23.6 & $-250 \pm 160 \hspace{0.4in}$\\ 
203     &       173     &       255     &       $5280 \pm 550$  &       31.6 & $140 \pm 120 \hspace{0.4in}$\\ 
289     &       261     &       342     &       $3660 \pm 340$  &       18.7 & $120 \pm 120 \hspace{0.4in}$\\ 
377     &       342     &       418     &       $1650 \pm 200$  &       7.3  & $160 \pm 140 \hspace{0.4in}$\\ 
465     &       418     &       500     &       $1890 \pm 220$  &       4.3  & $70 \pm 240 \hspace{0.4in}$\\ 
553     &       506     &       594     &       $2840 \pm 290$  &       4.0  & $0 \pm 300 \hspace{0.4in}$\\ 
641     &       600     &       676     &       $1670 \pm 270$  &       2.3  & $120 \pm 420 \hspace{0.4in}$\\ 
725     &       676     &       757     &       $2010 \pm 350$  &       1.7  & $-90 \pm 580 \hspace{0.4in}$\\ 
837     &       763     &       864     &       $2320 \pm 450$  &       1.1  & $-490 \pm 850 \hspace{0.4in}$\\

\hline\hline
\end{tabular}
\end{center}
The uncertainties listed above do not include flux scale and beam
calibration uncertainties. The total combined calibration uncertainty is 8\%
(1\,$\sigma$), expressed as a fractional uncertainty on the $C_l$ band
powers (4\% in $\Delta T/T$).
\tablenotetext{a}{$l_\mathrm{eff}$ is the band-power window function
weighted mean multipole moment (see text).}
\tablenotetext{b}{$l_\mathrm{l}$ and $l_\mathrm{h}$ are the low and
high $e^{-1/2}$ points of the band-power window function.}
\tablenotetext{c}{$\sigma_{\mathrm{s}}/\sigma_{\mathrm{n}}$ is the
ratio of the uncertainty attributable to sky sample variance to the
uncertainty attributable to noise (see text).}
\tablenotetext{d}{Power spectrum of the epoch-differenced data vector
described in \S\ref{sec:consistency}.}
\end{table*}

\begin{table*}[b]
\caption{\label{tab:corr}Correlation coefficient matrix for the DASI band powers given 
	in Table \ref{tab:ps}}
\begin{center}
\begin{tabular}{ccccccccc}
\hline\hline
$         1$    &$    -0.243$   &$    0.0349$   &$ -5.87\times 10^{-3}$  &$  1.09\times 10^{-3}$  &$ -3.01\times 10^{-4}$  &$ -1.08\times 10^{-4}$  &$ -1.24\times 10^{-4}$  &$ -1.99\times 10^{-4}$     \\ 
                &$         1$   &$    -0.182$   &$    0.0286$   &$ -6.65\times 10^{-3}$  &$  3.90\times 10^{-4}$  &$ -1.00\times 10^{-3}$  &$ -6.88\times 10^{-4}$  &$ -4.85\times 10^{-4}$     \\ 
                &               &$         1$   &$    -0.196$   &$    0.0372$   &$ -8.84\times 10^{-3}$  &$ -1.61\times 10^{-3}$  &$ -2.82\times 10^{-3}$  &$ -2.74\times 10^{-3}$     \\ 
                &               &               &$         1$   &$    -0.234$   &$    0.0334$   &$   -0.0149$   &$ -5.52\times 10^{-3}$  &$ -6.69\times 10^{-3}$     \\ 
                &               &               &               &$         1$   &$    -0.193$   &$    0.0247$   &$   -0.0193$   &$ -8.22\times 10^{-3}$     \\ 
                &               &               &               &               &$         1$   &$    -0.219$   &$    0.0394$   &$   -0.0200$      \\ 
                &               &               &               &               &               &$         1$   &$    -0.275$   &$    0.0339$      \\ 
                &               &               &               &               &               &               &$         1$   &$    -0.220$      \\ 
                &               &               &               &               &               &               &               &$         1$      \\ 

\hline\hline
\end{tabular}
\end{center}
\end{table*}

\subsection{Consistency Tests}
\label{sec:consistency}
We perform three types of tests to check the consistency of the data:
i) $\chi^2$ tests on the difference between two visibility data
vectors constructed from observations of the same fields on the sky,
ii) construction of a nine-band power spectrum of the
epoch-differenced visibility data vector, to test for significant
deviation from zero power, and iii) $\chi^2$ tests on the difference
between two power spectra constructed from independent fields on the
sky. In the second and third types of test, we increase the number of
frequency channels which are combined in the data vector in order to
reduce the computational time required to produce the power
spectra. This increased data compression yields a power
spectrum similar to the one reported above.

Of the three types of test, the first is the most powerful tool for
detecting non-Gaussianity or incorrect estimates of the noise. The
reduced $\chi^2$ statistic is 
\begin{equation}
\chi^2/N = (\Delta_1 - \Delta_2)^T (C_{n1} + C_{n2} + C_{Cg})^{-1} 
	(\Delta_1 - \Delta_2)/N,
\end{equation}
where $\Delta_1$ and $\Delta_2$ are the two data vectors, $C_{n1}$ and
$C_{n2}$ are the (diagonal) noise covariance matrices, $C_{Cg}$ is the
same ground constraint matrix that is used in the power spectrum
likelihood analysis, and $N = \mathrm{dim}\ \Delta - \mathrm{rank}\
C_{Cg}$ are the degrees of freedom. We split the visibilities between
the two epochs of available observations for each field row, yielding
$\chi^2/N = 1.03$. This $\chi^2$ value is significant given the $N
\sim 3 \times 10^4$ degrees of freedom --- it indicates that the noise
may be slightly non-Gaussian. In fact, we see improvement of this
statistic if we increase the severity of the lunar cuts, but the
effect on the power spectrum is negligible. It may also indicate that
we slightly underestimate the noise of the data.  However, the
uncertainties in all bands are dominated by sky sample variance,
rather than instrument noise, making the power spectrum robust against
a noise underestimate of this magnitude.

A power spectrum in nine bands was created from the epoch-differenced
data vector, and tested for deviation from zero power using a $\chi^2$
statistic, with the result $\chi^2/N = 9.5/9$, which is consistent
within the 68\% confidence interval. The band powers for the
epoch-differenced power spectrum are given in Table~\ref{tab:ps}.

We use a $\chi^2$ statistic to test the consistency between power
spectra generated from each of the four field rows. The $\chi^2$
statistic is constructed from the difference between two power spectra
which sample independent sky, $\chi^2 = (\mathcal{D}_1 -
\mathcal{D}_2)^T (P_1+P_2)^{-1}(\mathcal{D}_1 -
\mathcal{D}_2)$, where $\mathcal{D}_1,\mathcal{D}_2$ are
the band-power vectors and $P_1,P_2$ are the band-power covariance
(inverse Fisher) matrices.  The non-Gaussianity of the DASI power
spectrum uncertainties is small, which justifies using a $\chi^2$
statistic; we have tested its validity with Monte Carlo techniques on
simulated data and have not found a significant deviation from a
$\chi^2$ distribution. The resulting values, with format $\chi^2/N$
($\chi^2$ cumulative distribution function percentile), are: $14.9/9$
(91\%), $13.3/9$ (85\%), $10.9/9$ (72\%), $4.5/9$ (12\%), $6.2/9$
(28\%) and $3.7/9$ (7\%) for the A--B, A--C, A--D, B--C, B--D, C--D
differenced field row pairs, respectively. The power spectra of the four
field rows are in reasonable agreement.

To test the efficacy of the point source constraints described in \S
\ref{sec:analysis}, we split the data in each field row between the
four fields with the highest and four with the lowest
detected point source flux, and we create two power spectra from the two sets
of combined fields. The  $\chi^2/N$ value for the difference between the
two power spectra is $11.5/9$ (75\%) indicating they are consistent within
the 68\% confidence interval.

Although point sources are the foreground of primary concern for DASI,
constraint matrices are demonstrably effective in removing this point source power,
and the consistency tests above show that the power spectra from
sets of fields with very different point source flux contributions are in
good agreement after the constraint matrix is applied.

\subsection{Diffuse Foregrounds}
\label{sec:foregrounds}

Diffuse foregrounds are expected to be low in the region of the 
DASI fields at our observing frequency (see Paper I). To check this,
we place limits on the contribution of diffuse foregrounds to
the power spectrum by creating constraint matrices from foreground
templates. The constraint matrix formalism is a powerful technique for
placing limits on foregrounds with a known relative intensity distribution, since
it allows for arbitrary scaling of the template amplitude and spectral
index, without knowledge of these quantities at microwave frequencies. We
create foreground images centered on each of the DASI fields from the
cleaned IRAS 100~$\micron$ maps of \markcite{finkbeiner99}{Finkbeiner}, {Davis}, \&  {Schlegel} (1999), cleaned 408~MHz 
Haslam survey maps \markcite{haslam81,finkbeiner01}({Haslam} {et~al.} 1981; {Finkbeiner} 2001), and
H$\alpha$ maps \markcite{gaustad00,mccullough01}({Gaustad} {et~al.} 2000; {McCullough} 2001).  These images are
converted to visibility template vectors with the DASI simulation
software. We marginalize over modes in the data which match the
templates using the constraint matrix formalism described in \S
\ref{sec:analysis}. We constrain an arbitrary template amplitude and
spectral index for each DASI field. With the addition of all of these
foreground constraints, the maximum change in a band power is 3.3\%,
with most bands changing by less than 1\%. 

The Haslam map has a resolution of $\sim1\degr$, making it inadequate
as a template for multipole moments $\gtrsim 200$, however, the power
spectrum of synchrotron emission is expected to decrease with $l$
\markcite{tegmark96}(Tegmark \& Efstathiou 1996).  Also, the H$\alpha$ images show very low emission
in the region of the DASI fields, and are of questionable use as a
template. As a second method of characterizing possible free-free
emission, we convert the H$\alpha$ images to brightness temperature at
our frequencies assuming a gas temperature of $10^4$~K
\markcite{kulkarni87}({Kulkarni} \& {Heiles} 1988). Subsequent power spectrum analysis of the
resulting image visibility templates yields a $< 1$\% contribution to
our band powers in all bands. 

We conclude that dust, free-free, and
synchrotron emission, as well as emission with \emph{any} spectral
index that is correlated with these templates, such as spinning dust
grain emission \markcite{draine98}(Draine \& Lazarian 1998), make a negligible contribution to the
CMB power spectrum presented here.

\section{Conclusion}
\label{sec:conclusion}

In its initial season, the Degree Angular Scale Interferometer has
successfully measured the angular power spectrum of the CMB over the
range $l$ = 100--900 in nine bands with high precision. The
interferometer provides a simple and direct measurement of the power
spectrum, with systematic effects, calibration methods, and beam
uncertainties which are very different from single-dish
experiments. We have made extensive use of constraint matrices in the
analysis as a simple method for projecting out undesired signals in
the data, including ground-signal common modes and point sources with
arbitrary spectral index. The constraint matrix formalism is also used
as a powerful test of correlations with foreground templates with
arbitrary flux and spectral index scaling; we find no evidence of
diffuse foregrounds in the data.

We see strong evidence for both first and second peaks in the angular
power spectrum at $l \sim 200$ and $l \sim 550$, respectively, and a
rise in power at $l \sim 800$ that is suggestive of a third. The
detection of harmonic peaks in the power spectrum is a resounding
confirmation that sub-degree scale anisotropy in the CMB is the result 
of gravitationally driven acoustic oscillations such as
those which arise naturally in adiabatic inflationary theories. In
addition, the rise in power in the region of the predicted third peak
strongly supports, from CMB data alone, the presence of dark matter in
the Universe.

\acknowledgments

In this analysis, the authors were guided by the early efforts of
Martin White, and by numerous enlightening conversations with Lloyd
Knox, to whom we give our thanks. We are grateful for the efforts of
Stephan Meyer who, as the director of the Center for Astrophysical
Research in Antarctica (CARA), lent great support to the project. We
are indebted to our intrepid winterover crew, John Yamasaki and Gene
Davidson, for keeping the telescope in working order, and to Ethan
Schartman for his extensive hardware contributions. We thank Peter
McCullough and John Gaustad for providing preliminary data from the
H$\alpha$ Sky Survey. This research is supported by the National
Science Foundation under a cooperative agreement (NSF OPP 89-20223)
with CARA, a National Science Foundation Science and Technology
Center. Support at Caltech is provided by NSF grants AST 94-13935 and
AST 98-02989.

\bibliography{}


\begin{thebibliography}{}

\bibitem[Albrecht, Coulson, Ferreira, \&  Magueijo 1996]{albrecht96}
Albrecht, A., Coulson, D., Ferreira, P., {et al.}, 1996, Phys. Rev.  Lett., 76, 1413

\bibitem[Albrecht \& Steinhardt 1982]{albrecht82}
Albrecht, A. \& Steinhardt, P.~J. 1982, Phys. Rev. Lett., 48, 1220

\bibitem[{Bardeen}, {Steinhardt}, \&  {Turner} 1983]{bardeen83}
{Bardeen}, J.~M., {Steinhardt}, P.~J., \& {Turner}, M.~S. 1983, \prd, 28, 679

\bibitem[Bond \& Efstathiou 1984]{bond84}
Bond, J.~R. \& Efstathiou, G. 1984, \apjl, 285, L45

\bibitem[{Bond}, {Jaffe}, \& {Knox} 1998]{bond97}
{Bond}, J.~R., {Jaffe}, A.~H., \& {Knox}, L. 1998, \prd, 57, 2117

\bibitem[{Bond}, {Jaffe}, \& {Knox} 2000]{bond98}
---. 2000, \apj, 533, 19, astro-ph/9808264

\bibitem[Chamberlin, Lane, \& Stark 1997]{chamberlin97}
Chamberlin, R.~A., Lane, A.~P., \& Stark, A.~A. 1997, \apj, 476, 428

\bibitem[{de Bernardis}, Ade, Bock, Bond,  Borrill, Boscaleri, Coble, Crill, De~Gasperis, Farese, Ferreira, Ganga,  Giacometti, Hivon, Hristov, Iacoangeli, Jaffe, Lange, Martinis, Masi, Mason,  Mauskopf, Melchiorri, Miglio, Montroy, Netterfield, Pascale, Piacentini,  Pogosyan, Prunet, Rao, Romeo, Ruhl, Scaramuzzi, Sforna, \&  Vittorio 2000]{debernardis00}
{de Bernardis}, P., Ade, P. A.~R., Bock, J.~J., {et al.}, 2000, Nature,  404, 955

\bibitem[Draine \& Lazarian 1998]{draine98}
Draine, B.~T. \& Lazarian, A. 1998, \apj, 508, 157

\bibitem[{Finkbeiner} 2001]{finkbeiner01}
{Finkbeiner}, D.~P. 2001, private communication

\bibitem[{Finkbeiner}, {Davis}, \&  {Schlegel} 1999]{finkbeiner99}
{Finkbeiner}, D.~P., {Davis}, M., \& {Schlegel}, D.~J. 1999, \apj, 524, 867

\bibitem[{Gaustad}, {Rosing}, {McCullough}, \& {van  Buren} 2000]{gaustad00}
{Gaustad}, J.~E., {Rosing}, W., {McCullough}, P.~R., {et al.}, 2000,  \pasp, 220, 169

\bibitem[{Guth} 1981]{guth81}
{Guth}, A.~H. 1981, \prd, 23, 347

\bibitem[{Guth} \& {Pi} 1982]{guth82}
{Guth}, A.~H. \& {Pi}, S.~. 1982, Physical Review Letters, 49, 1110

\bibitem[Halverson \& Carlstrom 2001]{halverson01b}
Halverson, N. \& Carlstrom, J.~E. 2001, IEEE-MTT, to be submitted

\bibitem[{Hanany}, {Ade}, {Balbi}, {Bock}, {Borrill},  {Boscaleri}, {de Bernardis}, {Ferreira}, {Hristov}, {Jaffe}, {Lange}, {Lee},  {Mauskopf}, {Netterfield}, {Oh}, {Pascale}, {Rabii}, {Richards}, {Smoot},  {Stompor}, {Winant}, \& {Wu} 2000]{hanany00}
{Hanany}, S., {Ade}, P., {Balbi}, A., {et al.}, 2000, \apjl, 545, L5, astro-ph/0005123

\bibitem[{Haslam}, {Klein}, {Salter}, {Stoffel},  {Wilson}, {Cleary}, {Cooke}, \& {Thomasson} 1981]{haslam81}
{Haslam}, C. G.~T., {Klein}, U., {Salter}, C.~J., {et al.}, 1981, \aap, 100,  209

\bibitem[{Hawking} 1982]{hawking82}
{Hawking}, S.~W. 1982, Phys.\ Lett.\ B, 115, 295

\bibitem[Hobson, Lasenby, \& Jones 1995]{hobson95}
Hobson, M.~P., Lasenby, A.~N., \& Jones, M. 1995, \mnras, 275, 863

\bibitem[Hu \& White 1996]{hu96a}
Hu, W. \& White, M. 1996, \apj, 471, 30

\bibitem[{Jones} 1996]{jones96}
{Jones}, M.~E. 1996, in Moriond Astrophysics Meetings, Vol. XVI, Microwave  Background Anistropies, ed. B.~G. J.~V. F.R.~Bouchet, R.~Gispert  (Gif-sur-Yvette: Editions Frontieres), 161, {ISBN}: 3863322087

\bibitem[{Jungman}, {Kamionkowski}, {Kosowsky}, \&  {Spergel} 1996]{jungman96}
{Jungman}, G., {Kamionkowski}, M., {Kosowsky}, A., {et al.}, 1996,  \prd, 54, 1332

\bibitem[{Knox} 1995]{knox95}
{Knox}, L. 1995, \prd, 52, 4307

\bibitem[{Knox} 1999]{knox99}
---. 1999, \prd, 60, 103516, astro-ph/9902046

\bibitem[{Kulkarni} \& {Heiles} 1988]{kulkarni87}
{Kulkarni}, S.~R. \& {Heiles}, C. 1988, in Galactic and Extragalactic Radio  Astronomy, Second Edition, ed. G.~L. Verschuur \& K.~I. Kellerman (New York:  Springer-Verlag), 95, {ISBN}: 0-387-96575-0

\bibitem[{Lay} \& {Halverson} 2000]{lay98}
{Lay}, O.~P. \& {Halverson}, N.~W. 2000, \apj, 543, 787

\bibitem[Leitch, Pryke, Halverson, Carlstrom, Kovac,  Holzapfel, Dragovan, Cartwright, Mason, Padin, Pearson, Readhead, \&  Shepherd 2001]{leitch01}
Leitch, E.~M., Pryke, C., Halverson, N.~W., {et al.}, 2001, \apj, 568, 28,  astro-ph/0104488

\bibitem[Linde 1982]{linde82}
Linde, A.~D. 1982, Phys. Lett., B108, 389

\bibitem[{Mauskopf}, {Ade}, {de Bernardis}, {Bock},  {Borrill}, {Boscaleri}, {Crill}, {DeGasperis}, {De Troia}, {Farese},  {Ferreira}, {Ganga}, {Giacometti}, {Hanany}, {Hristov}, {Iacoangeli},  {Jaffe}, {Lange}, {Lee}, {Masi}, {Melchiorri}, {Melchiorri}, {Miglio},  {Montroy}, {Netterfield}, {Pascale}, {Piacentini}, {Richards}, {Romeo},  {Ruhl}, {Scannapieco}, {Scaramuzzi}, {Stompor}, \& {Vittorio} 2000]{mauskopf00a}
{Mauskopf}, P.~D., {Ade}, P.~A.~R., {de Bernardis}, P., {et al.}, 2000, \apjl, 536, L59

\bibitem[{McCullough} 2001]{mccullough01}
{McCullough}, P.~R. 2001, private communication

\bibitem[{Miller}, {Caldwell}, {Devlin}, {Dorwart},  {Herbig}, {Nolta}, {Page}, {Puchalla}, {Torbet}, \& {Tran} 1999]{miller99}
{Miller}, A.~D., {Caldwell}, R., {Devlin}, M.~J., {et al.}, 1999, \apjl, 524, L1

\bibitem[{Padin}, {Cartwright}, {Mason},  {Readhead}, {Shepherd}, {Sievers}, {Udomprasert}, {Holzapfel}, {Myers},  {Carlstrom}, {Leitch}, {Joy}, {Bronfman}, \& {May} 2001a]{padin01}
{Padin}, S., {Cartwright}, J.~K., {Mason}, B.~S., {et al.}, 2001a, \apj, 549, L1, astro-ph/0012211

\bibitem[{Padin}, {Cartwright},  {Shepherd}, {Yamasaki}, \& {Holzapfel} 2001b]{padin00}
{Padin}, S., {Cartwright}, J.~K., {Shepherd}, M.~C., {et al.}, 2001b, IEEE Trans. Instrum. Meas., 50, 1234

\bibitem[{Pearson}, {Readhead}, {Padin}, {Cartwright},  {Mason}, {Myers}, {Shepherd}, {Sievers}, \& {Udomprasert} 2000]{pearson00}
{Pearson}, T.~J., {Readhead}, A.~C.~S., {Padin}, S., {et al.}, 2000, in IAU Symp. Proc., Vol. 201, New Cosmological  Data and the Values of the Fundamental Parameters, ed. A.~Lasenby \&  A.~Wilkinson (ASP), astro-ph/0012212

\bibitem[Pospiezalski, Lakatosh, Nguyen, Lui, Liu,  Le, Thompson, \& Delaney 1995]{pospieszalski95}
Pospiezalski, M.~W., Lakatosh, W.~J., Nguyen, L.~D., {et al.}, 1995, IEEE MTT-S Int. Microwave Symp.,  1121

\bibitem[Pryke, Halverson, Leitch, Carlstrom, Kovac,  Holzapfel, \& Dragovan 2001]{pryke01}
Pryke, C., Halverson, N.~W., Leitch, E.~M., {et al.}, 2001, \apj, 568, 46, astro-ph/0104490

\bibitem[Smoot {et~al.} 1992]{smoot92}
Smoot, G.~F. {et~al.} 1992, \apj, 396, L1

\bibitem[Starobinsky 1982]{starobinskii82}
Starobinsky, A.~A. 1982, Phys. Lett., B117, 175

\bibitem[Tegmark \& Efstathiou 1996]{tegmark96}
Tegmark, M. \& Efstathiou, G. 1996, \mnras, 281, 1297

\bibitem[{Vittorio} \& {Silk} 1984]{vittorio84}
{Vittorio}, N. \& {Silk}, J. 1984, \apjl, 285, L39

\bibitem[White, Carlstrom, Dragovan, \&  Holzapfel 1999a]{white99a}
White, M., Carlstrom, J.~E., Dragovan, M., {et al.}, 1999a, \apj, 514, 12, astro-ph/9712195

\bibitem[White, Carlstrom, Dragovan,  Holzapfel, Halverson, Kovac, \& Leitch 1999b]{white99b}
White, M., Carlstrom, J.~E., Dragovan, M., {et al.}, 1999b, exists only as  astro-ph/9912422

\bibitem[White, Scott, \& Silk 1994]{white94}
White, M., Scott, D., \& Silk, J. 1994, \araa, 32, 319

\bibitem[{Wright}, {Griffith}, {Burke}, \&  {Ekers} 1994]{wright94}
{Wright}, A.~E., {Griffith}, M.~R., {Burke}, B.~F., {et al.}, 1994,  \apjs, 91, 111

\end{thebibliography}

\end{document}